\documentstyle[aps,preprint,tighten,psfig]{revtex}

\newcommand{\beq}{\begin{equation}}
\newcommand{\eeq}[1]{\label{#1} \end{equation}}
\newcommand{\beqar}{\begin{eqnarray}}
\newcommand{\eeqar}[1]{\label{#1} \end{eqnarray}}

\begin{document}
\draft
\preprint{CU-TP-898}

\title{Baryon Number Transport via Gluonic Junctions}

\author{Stephen E. Vance, Miklos Gyulassy}
\address{Physics Department, Columbia University, New York, N.Y. 10027}

\author{Xin-Nian Wang}
\address{Nuclear Science Division, Lawrence Berkeley National Laboratory
(LBNL), Berkeley, CA 94720}

\date{8 June 1998}

\maketitle

\begin{abstract} 
A novel non-perturbative gluon junction mechanism
is introduced within the HIJING/B nuclear collision event
generator to calculate baryon number transport and hyperon production
in pA and AA collisions.
This gluonic mechanism can account for the observed large
mid-rapidity valence baryon yield in $Pb+Pb$ at 160 AGeV
 and predicts high initial baryon densities at RHIC.  
However, the highly enhanced $\Lambda-\bar{\Lambda}$ yield 
and the baryon transverse momentum flow observed in this reaction
can only be partially described.
\end{abstract}
\pacs{}

\narrowtext

Recent data
\cite{na35_hyperons,na35_charged,na44_protons,na44_ppbar,na49_qm97} 
on $p+A$ and $A+B$ interactions at the CERN SPS has revealed 
a large degree of stopping and strange hyperon production in 
the heavy nuclear systems.  The stopping is significantly
under-predicted by models which assume that the primary mechanism
for baryon transport is 
diquark-quark ($qq-q$) hadronic strings \cite{gyu_stp97,topor_95}.  
In this letter we implement 
a variant\cite{khar_bj96} of the baryon junction 
mechanism\cite{rossi_77,kop_89} to address this problem.
This implementation is now available as a Monte Carlo event generator,
HIJING/B\cite{vance_hijb}, a new version of the HIJING model\cite{hijing}.  

In Fig. 1, the rapidity distributions of valence protons and lambdas
are compared to HIJING predictions.  It is clear that even in $p+S$
reactions, the mid-rapidity baryon distribution is largely under-predicted.
In HIJING, as in the LUND Fritiof \cite{fritiof} and 
dual parton model (DPM) \cite{dpm_94}, the valence baryons cannot move inward
from the beam and target fragmentation regions by more than about 2 units of
rapidity because of the assumed diquark fragmentation dynamics of 
the $qq-q$ strings.

Several models have considered additional mechanisms or string 
configurations to correct this problem.   
In the VENUS model\cite{venus}, the stopping and
strange baryon production is reproduced by adding an {\em ad hoc} double 
string mechanism whose parameters are adjusted to fit the data.
The RQMD\cite{rqmd} and UrQMD\cite{urqmd} event generators reproduce
the observed stopping through incoherent multiple inelastic 
scattering of the valence (di)quarks.    
The difficulty with this idea is 
that the time dilation of inelastic processes at high energies leads to 
coherence which is absent in this model.

In a new version of the DPM model\cite{dpm_dqb}, enhanced baryon stopping 
is more naturally achieved by introducing a diquark breaking 
mechanism\cite{kop_89}.  In this approach, a gluon exchange mediates 
the break-up of the diquark by 
changing its color state from a $\{\bar{3}\}$ to a $\{6\}$.   
The cross section for this interaction is then estimated \cite{kop_89} to be 
\beq
\frac{d\sigma}{dy} ( p h \rightarrow p X ) \approx  C \left ( \frac{1 GeV}
{\sqrt{s}} \right )^{1/2} \cosh(y/2),
\eeq{kop}
where $C \simeq 7$ mb was estimated to fit the observed baryon stopping
in pp collisions at the ISR \cite{kop_89}.   
The unusual $\cosh(y/2)$ rapidity dependence
and $1/\sqrt[4]{s}$ energy dependence 
follow here from the Regge motivated $1/\sqrt{x}$ distribution of the 
valence quarks in the proton at small x.   
The resulting configuration is a $q-q-q$ string where the 
baryon number is associated with the middle quark.  
In nucleus-nucleus collisions, 
diquarks are assumed to remain broken once they transform into a $\{6\}$
color state.
Thus, in $A+B$ collisions, the probability for diquark breaking should
be $P^{AB}_{DB} = 1-(1-P^{NN}_{DB})^n$, where $n$ is the number 
of collisions and $P^{NN}_{DB}$ is the diquark breakup probability
in a nucleon-nucleon collision. For a heavy nucleus, very few of the diquarks
survive.  The addition of this diquark breaking 
component was shown in \cite{dpm_dqb} to reproduce 
the observed stopping in AA collisions at the SPS.
However, this mechanism alone could not account for the observed 
strangeness enhancement without including additional final state 
interaction effects
\cite{dpm_strangeness}. 

In this letter, we consider another Regge motivated mechanism\cite{rossi_77}
which also leads to eq. (\ref{kop}) but which differs from the diquark breaking
model in multiplicity and strangeness enhancements.   
This mechanism is motivated from the non-perturbative gluon field 
configuration (the baryon junction) that appears
when writing the simplest gauge
invariant operator for the baryon in ${\rm SU}_c(3)$;
\beqar
B =  &\epsilon^{j_1j_2j_3} &
\left [ P \exp \left (ig \int_{x_1}^{x_J} dx^{\mu} A_{\mu} \right ) q(x_1)
\right ]_{j_1} 
\left [P \exp \left (ig \int_{x_2}^{x_J} dx^{\mu} A_{\mu} \right ) q(x_2) 
\right ]_{j_2} \nonumber \\  & \times & 
\left [P \exp \left (ig \int_{x_3}^{x_J} dx^{\mu} A_{\mu} \right ) q(x_3) 
\right ]_{j_3}. 
\eeqar{baryon}
Here, the baryon junction is the vertex at $x_J$ where the three gluon        
Wilson lines link the three valence quarks
to form the gauge invariant non-local operator.
In a highly excited baryonic state, the Wilson lines represent
color flux tubes.   When these strings fragment via $q\bar{q}$ production, 
the resulting baryon will be composed of the three sea quarks   
which are linked to the junction while the original valence quarks will 
emerge as constituents of three leading mesons.  Being a gluonic 
configuration, it was proposed\cite{khar_bj96} that the junction 
could be more easily
transported into the mid-rapidity region in hadronic interactions.

In Regge phenomenology, allowing for the possibility of baryon junction
exchanges in $pp$ and $p\bar{p}$ scattering is taken into account by 
adding new Regge trajectories ($M^J$).  Here, the exchange of the leading
$M^J_0$ trajectory is characterized by a three jet event 
in contrast to the two jet event resulting from the Pomeron exchange. 
Phenomenologically, the difference between the $p\bar{p}$ 
and $pp$ topological cross sections, 
$\Delta \sigma = \sigma(p\bar{p}) - \sigma(pp)$, 
is given by the contributions of the $\rho$, 
$\omega$ and $M^J_0$ Regge exchanges\cite{rossi_77}.  
Analysis of the energy dependence and the multiplicity distributions 
of $pp$ and $p\bar{p}$ data shows that $\Delta \sigma 
\propto s^{-1/2}$\cite{camilleri_87}
and that the difference, $\Delta \sigma$, may be largely 
associated with annihilations\cite{miettinen_76}.
Upon associating $\Delta \sigma$ with the annihilations 
or the $M^J$ exchanges,  the intercept of the leading term, 
$\alpha_{M^J_0}(0)$, can be estimated from the energy dependence 
of $\Delta \sigma$, i.e.
$\Delta \sigma \propto s^{-1/2} \simeq s^{1 - \alpha_{M^J_0}(0)}$, 
giving $\alpha_{M^J_0}(0) \simeq 1/2$.  
This value of $\alpha_{M^J_0}(0) \simeq 1/2$ is also consistent 
with estimates of 
the intercept using multi-peripheral model 
approximations\cite{eylon_74,rossi_77}.   
In addition, various models which associated $\Delta \sigma$ 
with three sheet, $M^J_0$ type annihilation events were able to 
explain the observed mean multiplicity, $\bar{n}_{B\bar{B}ann} 
\simeq \frac{3}{2} \bar{n}_{B\bar{B}scatt}$\cite{rossi_77}, the 
observed multiplicity distributions of $\Delta \sigma$\cite{webber_76}, 
and the experimental values of the ratio 
$R_n = (\sigma(p\bar{p})_n - \sigma(pp)_n )/\sigma(pp)_n$,
where $n$ is the prong number\cite{camilleri_87}. 
Recently, Kharzeev\cite{khar_bj96} used 
the exchange of the $M^J_0$ Reggeon in baryon production to propose 
a new baryon stopping mechanism.  This mechanism is  
able account for the excess 
of mid-rapidity valence baryons in $pp$ at ISR energies and was 
suggested to provide substantial stopping at RHIC and the LHC.  
In this letter, we implement this new baryon stopping mechanism into 
the Monte Carlo event generator HIJING/B, 
and tests its implications for RHIC.

As was shown by Kharzeev\cite{khar_bj96}, the energy and rapidity dependence 
of the inclusive baryon production at mid-rapidity can be obtained 
using Mueller's generalized optical theorem \cite{mueller_70} in the 
double Regge limit.  Here, the exchanges of a Pomeron and 
a $M^J_0$ Reggion lead to the following form for single mid-rapidity baryon
production,
\beq
E_B \frac{d^3\sigma^{(1)}}{d^3 p_B} = C_B f_B(m^2_t) \left ( \frac{s_0}{s}
 \right )^{1/4} \cosh(y/2)\;\; .
\eeq{bj1}
where $C_B$ is a constant that reflects the couplings
of the Reggeon and Pomeron to the proton,
$f_B (m^2_t)$ is an unknown function of $m_t$ and 
$s_0 \simeq 1$ GeV is a Regge energy scale.
The $\cosh(y/2)$ rapidity dependence 
and the $1/\sqrt[4]{s}$ 
energy dependence follow from the assumed intercept, 
$\alpha_{M^J_0}(0) \approx 1/2$.  
In contrast to the diquark breaking model in the DPM, 
the multiplicity\cite{khar_bj96}
of these events is enhanced by a factor of 5/4 while the 
strangeness content is enhanced by a factor of 3.  
The enhancement factor of 3 of the strangeness content of the baryon allows 
for the unique possibility of producing $S=-3$ $\Omega^-$ baryons.

We note that the value of the junction intercept, $\alpha_{M^J_0}(0) = 1/2$, 
has been criticized by the authors of the diquark 
breaking mechanism\cite{kop_88}.
Following the phenomenology which motivated the diquark breaking
mechanism, they show that the exchange of two gluons in 
the $\{10\}$ color state (Decameron) leads to baryon number transport with
baryon junctions.  However, in contrast to the above model using 
the exchange of $M^J$ Reggeons, their model leads to a constant, 
energy independent annihilation cross section\cite{kop_88}, 
$\sigma_{\{10\}} \simeq 1 - 2 \;{\rm mb}$,
and a uniform, rapidity independent, inclusive cross section 
for mid-rapidity baryon production\cite{kop_89}, 
$d\sigma(pp \rightarrow pX)/dy \simeq 0.1 \;{\rm mb}$.    
They also fit\cite{kop_88} the multiplicity distributions of $\Delta \sigma$ 
at $E_{lab} = 10, 20, 50 \; {\rm and} \; 100 \; {\rm GeV}$  
to argue that $\Delta \sigma$ is dominated by 
the $\rho$ and $\omega$ exchanges, while the three sheet 
annihilation component only contributes a 
small energy independent cross section of $1 - 2 \;{\rm mb}$.  
The energy dependence of this small three sheet annihilation cross 
section would then imply that $\alpha_{M^J_0}(0) \simeq 1$.   

In HIJING/B\cite{vance_hijb}, we implement Kharzeev's\cite{khar_bj96} model 
through a ``Y'' string configuration for the excited baryon.
For reactions without junction exchange,
standard $qq-q$ strings are used. 
The baryon is resolved around the junction via $q \bar{q}$ production 
and the resulting three beam jets are fragmented as $q - \bar{q}$ strings.
The hard processes are modeled as kinks in one of the three 
$q\bar{q}$ strings.  A value of $\sigma_{BJ} = 18$ mb is taken to reproduce
valence proton data \cite{pp400} from $p+p$ collisions at 400 GeV/c 
incident momentum $(\sqrt{s_0} = 27.4\;{\rm GeV})$.  
In order for the three beam jets to have 
sufficient phase space to decay, the junction exchange is only allowed 
if the invariant mass of the excited ``Y'' configuration exceeds 
$m \ge 5$ GeV.  At SPS energies, this kinematic constraint
considerably limits the number of junction exchanges allowed, 
reducing its effective cross section to $\sim 9$ mb.  
Like the modified DPM, 
we assume for multiple collisions that the baryon junction remains 
stopped in subsequent soft interactions once it has been exchanged.
The $\vec{p}_T$ of the junction baryon is 
obtained by adding the $\vec{p}_T$ of the three sea quarks.
We note that the present version of HIJING/B does not include 
final state interactions,
as our main interest is to test the extent to which initial state
non-equilibrium dynamics can account for the observed
strangeness and $p_\perp$ enhancements.

In Fig 1, the valence proton rapidity $(dN/dy_p - dN/dy_{\bar{p}})$ 
and valence hyperon rapidity $(dN/dy_{\Lambda} - dN/dy_{\bar{\Lambda}})$ 
distributions of HIJING and  HIJING/B are compared with minimum bias $p+S$ 
data\cite{na35_hyperons,na35_charged} at 200 AGeV and central 
$Pb+Pb$ (published and preliminary) data at 158 AGeV 
\cite{na44_protons,na49_qm97}.  While 
HIJING severely under-predicts both the observed stopping and 
the hyperon production,  the junction physics of HIJING/B 
sufficiently enhances both the baryon stopping and the hyperon production at 
mid-rapidity, reproducing the $p+S$ results.  However, as seen in Fig 1d, 
even with the sizeable strangeness enhancement factor of 3, 
this gluonic mechanism can only partially account for the 
observed hyperon enhancement in the heavy systems such as $Pb+Pb$.

The $p_T$ distributions of the protons are calculated in HIJING/B
for $p+Pb$ at beam momenta of 450 GeV/c and $S+Pb$ at beam momenta 
of 200 GeV/c in the rapidity interval, 
$2.3 < y < 2.9$.  The $m_T$ distributions were then fit 
for $m_T - m_p < 0.68$ GeV with the functional form 
$E d^3N/dp^3 = A \exp(-m_T/T)$, where $A$ is a 
normalization constant and $T$ is the inverse slope.
The inverse slopes were found to be $T \approx 151$ MeV for $p+Pb$ 
and $T \approx 168$ MeV for $S+Pb$.   The measured values from the NA44 
collaboration\cite{na44_ppbar} are $T = 195 \pm 5$ MeV for $p+Pb$ 
and $T = 256 \pm 4$ MeV for $S+Pb$.  
Although the $<p_T^2>$ of the junction baryons is enhanced by the 
a factor of 3 over the normal $qq-q$ string, the contribution from this mechanism alone is insufficient 
to reproduce the observed enhanced flow.  This results from the competition 
between the different $p_T$ distributions which arise from the different 
production mechanisms.   
If one-half of the protons emerge from $qq-q$ 
strings where $T_{dq} \approx$ 130 MeV, while the other one-half come from  
junction configurations where $T_{BJ} = \sqrt{3} T_{dq} \approx 225$ MeV, 
the effective inverse slope is only  $T_{eff} \approx 173$ MeV in the measured
region. The inability of this rather strong
initial state non-equilibrium dynamical mechanism along with the 
already included Cronin effect to account for the observed transverse 
baryon flow provides evidence for
its possible origin as due to final state interactions. 
We note also that the above junction dynamics alone provides no mechanism 
to  account  for  the observed enhancement of the anti-proton
$p_T$ \cite{na44_ppbar}.

The impact parameter dependence of this stopping mechanism is studied 
in Fig 2 for the net valence baryons, $B - \bar{B} = (p - \bar{p}) + 
(n - \bar{n}) + (\Lambda - \bar{\Lambda})$.   
For the impact parameters $b = 0 - 3$ fm, $b = 4 - 5$ fm and $b = 7 - 8$ fm,
a strong to moderate degree of baryon stopping is observed.  
However, at $b = 10 - 11$ fm, the degree of stopping has decreased and
the shape  suggests semi-transparency.
Recent data\cite{gonin_qm96} has shown the suppression of
$J/\Psi$ for impact parameter of $b \le 8$ fm.
Measurements of the impact parameter ($E_T$) dependence of 
baryon stopping would be of interest
to test if the anomalous $J/\psi$ suppression\cite{gonin_qm96}
threshold at $b\sim 8$ fm is correlated with the onset of greater baryon
stopping.   In the  baryon junction exchange picture, a large
degree of stopping is directly correlated with an
enhanced  gluonic field intensity at mid-rapidity that could partially be the
cause of the ionization of $c-\bar{c}$ pairs.

The predictions of this model for the valence proton and lambda rapidity 
distributions in $Au+Au$ collisions at 
RHIC energies ($\sqrt{s}$ = 200 GeV) are shown in Fig 3.
HIJING/B predicts approximately 
twice the initial number of valence protons and five 
times the initial number of valence hyperons of HIJING at mid-rapidity 
leading to a prediction of twice the initial baryon density, 
$\rho(\tau_0) \approx 2 \rho_0 \approx 0.3/{\rm fm}^3$.  
Previous predictions for RHIC assuming idealized
zero baryon chemical potential scenarios should therefore be re-examined.

\section{Acknowledgments}
We would like to thank D. Kharzeev, P. Jacobs and B.Z. Kopeliovic 
for stimulating and critical discussions.

This work was supported by the Director, Office of Energy Research,
Division of Nuclear Physics of the Office of High Energy and
Nuclear Physics of the U.S. Department of Energy under Contract No.
DE-FG02-93ER40764 and DE-AC03-76SF00098.

\newpage
\begin{figure}[htb]
\hspace{0.4cm}
\psfig{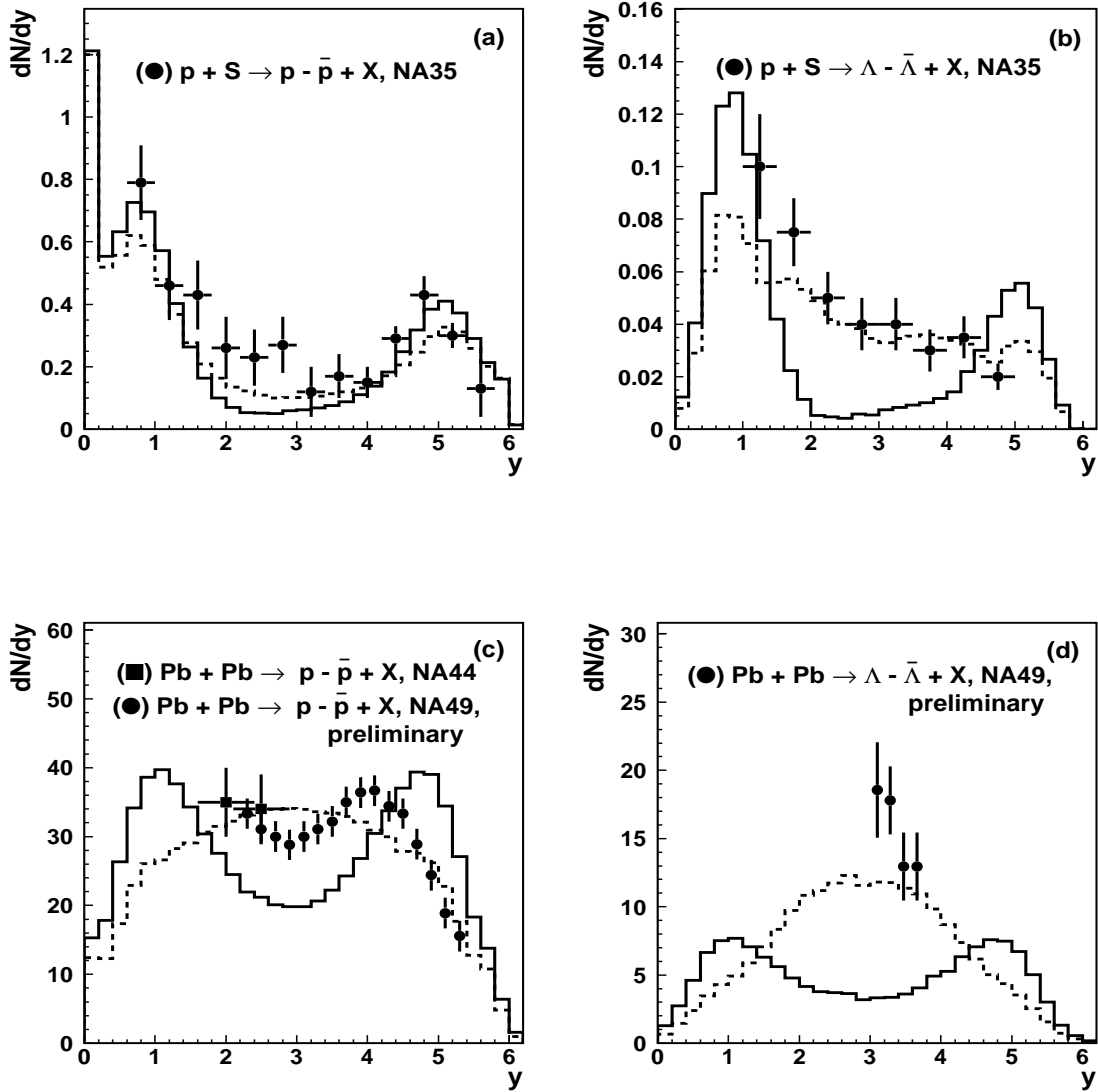}
\caption{HIJING (solid) and HIJING/B (dashed) calculations of the 
valence proton and hyperon rapidity distributions are shown for 
minimum bias $p+S$ collisions at 200 AGeV and central 
$Pb+Pb$ collisions at 160 AGeV.   
The data are from measurements made by the NA35
\protect \cite{na35_hyperons,na35_charged}, NA44
\protect \cite{na44_protons} and NA49 \protect \cite{na49_qm97} 
collaborations.}

\end{figure}

\newpage
\begin{figure}[htb]
\hspace{0.5in}
\psfig{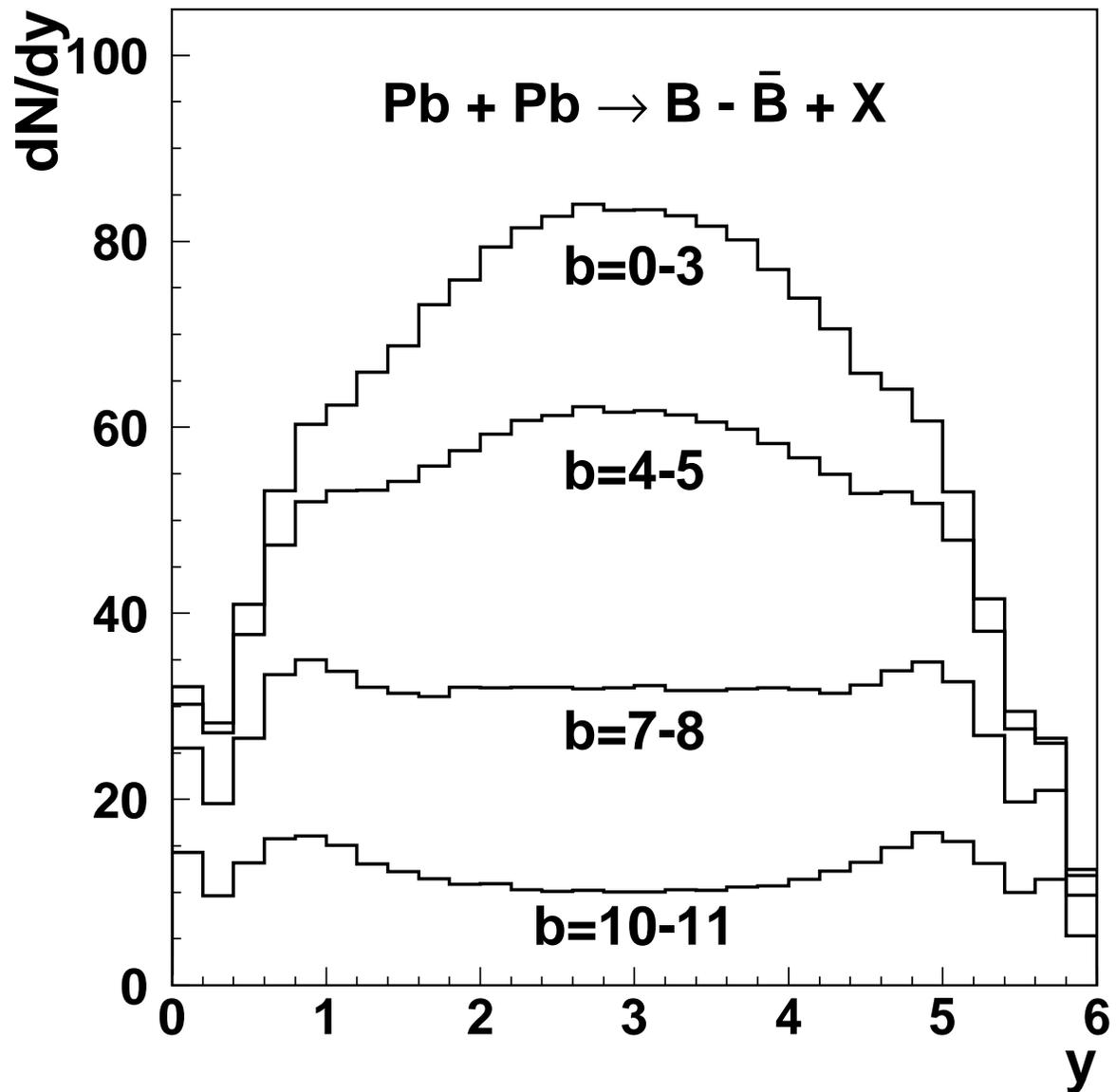}
\caption{Calculations of the net valence baryon rapidity distributions 
using HIJING/B are given for the impact parameter
windows of $b=0-3$ fm, $b=4-5$ fm, $b=7-8$ fm and $b=10-11$ fm.  
In this calculation, the net valence baryons are defined as $B - \bar{B} = 
(p - \bar{p}) + (n - \bar{n}) + (\Lambda - \bar{\Lambda})$. }
\end{figure}

\newpage
\begin{figure}[htb]
\psfig{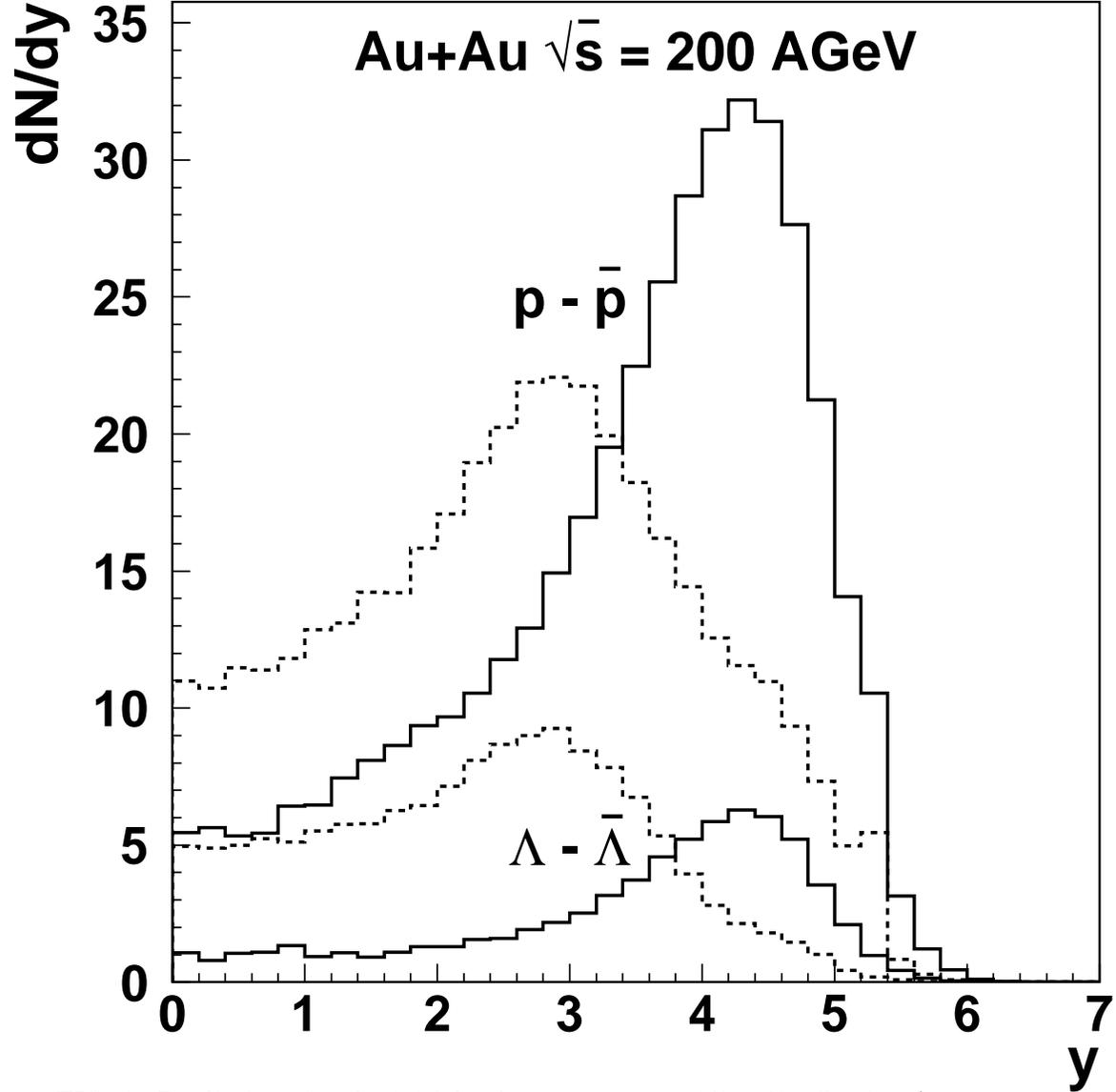}
\caption{Predictions for the initial valence proton rapidity 
distribution (upper two curves) and for the initial valence hyperon 
rapidity distribution (lower two curves) are given for 
Au+Au collisions at $E_{cm} = 200$ 
AGeV by HIJING (solid) and HIJING/B(dashed).}
\end{figure}


\begin{thebibliography}{99}

\bibitem{na35_hyperons} 
T. Alber et. al., (NA35 Collaboration), Z. Phys. {\bf C64} (1994) 195.

\bibitem{na35_charged}
T. Alber et. al., (NA35 Collaboration), Eur. Phys. J. {\bf C2} (1998) 643.

\bibitem{na44_protons} 
I.G. Bearden et al., (NA44 Collaboration), Phys. Lett. {\bf B388} (1996) 431. 

\bibitem{na44_ppbar}
K. Wolf et al., (NA44 Collaboration), Phys. Rev. {\bf C57} (1998) 837.

\bibitem{na49_qm97}
G. Roland et al., (NA49 Collaboration), Proceedings of the Thirteenth
International Conference on Ultra-Relativistic Nucleus-Nucleus Collisions,
Quark Matter '97. 

\bibitem{gyu_stp97} 
M. Gyulassy, V. Topor Pop, and S.E. Vance, Heavy Ion Physics
{\bf 5} (1997) 299.

\bibitem{topor_95} 
V. Topor Pop, et al., Phys. Rev. {\bf C52} (1995) 1618;  
M. Gyulassy, V. Topor Pop, X. N. Wang,  Phys. Rev. {\bf C54} (1996) 1497.

\bibitem{khar_bj96} 
D. Kharzeev, Phys. Lett. {\bf B378} (1996) 238. nucl-th/9602027.

\bibitem{rossi_77} 
G.C. Rossi and G. Veneziano, Nucl.Phys. B123 (1977) 507; 
Phys. Rep. {\bf 63} (1980) 153.

\bibitem{kop_89}
B.Z. Kopeliovich and B.G. Zakharov, Z. Phys. C, {\bf 43}, (1989) 241. 

\bibitem{vance_hijb}
HIJING/B available upon request from
svance@nt3.phys.columbia.edu.

\bibitem{hijing} 
X. N. Wang and M. Gyulassy, Phys. Rev. {\bf D44} (1991) 3501; 
Phys. Rev. {\bf D45} (1992) 844;
Comp. Phys. Comm. {\bf 83} (1994) 307 .

\bibitem{fritiof} 
B. Andersson, et al., Nucl. Phys. {\bf B281} (1987) 289;
Comp. Phys. Commun. {\bf 43} (1987) 387.

\bibitem{dpm_94} 
A. Capella, U. Sukhatme, C. I. Tan and
J. Tran Thanh Van, Phys. Rep. {\bf 236} (1994) 225.

\bibitem{venus}
K. Werner, Phys. Rep. {\bf 232} (1993) 87.

\bibitem{rqmd}
H. Sorge, Phys. Rev. {\bf C52} (1995) 3291; nucl-th/9509007.

\bibitem{urqmd}
S.A. Bass, et al.,  Prog. Part. Nucl. Phys. {\bf 41} (1998) 225;
nucl-th/9803035.

\bibitem{dpm_dqb}
A. Capella, B.Z. Kopeliovich,  Phys. Lett. {\bf B381} (1996) 325;
hep-ph/9603279. 

\bibitem{dpm_strangeness}
A. Capella,  Phys. Lett. {\bf B387} (1996) 400; hep-ph/9605216.

\bibitem{camilleri_87}
L. Camilleri, Phys. Rep. {\bf 144}, (1987) 51.  

\bibitem{miettinen_76}
H.I. Miettinen, Rapporteur's talk, 3rd European Symposium on
Antinucleon-Nucleon Reactions, Stockholm, 1976.

\bibitem{eylon_74}
Y. Eylon and H. Harari,  Nucl. Phys. {\bf B80} (1974) 349.


\bibitem{webber_76}
B.R. Webber, Nucl. Phys. {\bf B117} (1976) 445. 






\bibitem{mueller_70}
A.H. Mueller, Phys. Rev. {\bf D2} (1970) 2963.

\bibitem{kop_88}
B.Z. Kopeliovich and B.G. Zakharov,  Phys. Lett. {\bf B211} (1988) 221.



\bibitem{pp400} 
M. Aguilar-Benitez et al. (LEBC-EHS Collaboration), 
Z. Phys. {\bf C50} (1991) 405.

\bibitem{gonin_qm96}
M. Gonin (NA50 Collaboration), Proc. of the Quark Matter '96 Conf., Eds.  
P. Braun-Munzinger et al., Nucl. Phys. {\bf A610} (1996) 404c.   
\end{thebibliography}
\end{document}